\documentclass[aps,pra,floatfix,superscriptaddress,showpacs,twocolumn]{revtex4}
\usepackage{amsmath}
\usepackage{graphicx}

\begin{document}
\title{Rubidium ``whiskers'' in a vapor cell}

\author{M. V. Balabas}
\email{mbalabas@yandex.ru} \affiliation{S. I. Vavilov State Optical
Institute, St. Petersburg, 199034 Russia}
\author{Alexander O. Sushkov}
\email{alex000@socrates.berkeley.edu} \affiliation{Department of
Physics, University of California, Berkeley, CA 94720-7300}
\author{Dmitry Budker}
\email{budker@berkeley.edu} \affiliation{Department of Physics,
University of California, Berkeley, CA 94720-7300}
\affiliation{Nuclear Science Division, Lawrence Berkeley National
Laboratory, Berkeley CA 94720}

\date{\today}
\begin{abstract}
Crystals of metallic rubidium are observed ``growing'' from paraffin
coating of buffer-gas-free glass vapor cells. The crystals have
uniform square cross-section, $\approx 30\ \mu$m on the side, and
reach several mm in length.
\end{abstract}
\maketitle

For many years, we have been using glass alkali-metal vapor cells
with inner walls coated with paraffin. High-quality paraffin
coatings \cite{Rob58,BouBro66} have the ``magical'' property that
polarized alkali atoms may bounce between the cell walls several
thousand times before they depolarize, which has led to wide-spread
application of such cells anywhere where long-lived atomic
polarization is desired, for example, in optical atomic
magnetometers \cite{BudRom2006}.

While condensation of the alkali metal on the paraffin coated walls
is usually carefully avoided, and the excess metal is located in the
cell ``stem'' (see Fig. 1), in some of the cells, we have observed
needle-like metal crystals shown on the insets of Fig. 1, that have
``grown'' from the paraffin coating and have reached the length of
up to 2-3 mm in about a two-month's time since the cells were made
(our coating procedure is described in Ref. \cite{AleLIAD}). The
crystals are all of regular square cross-section, $\approx 30\ \mu$m
on the side, terminating on a sharp-vertex pyramid at the ends of
the needle.

The observed rate of growth indicates that an atom from the vapor
phase that hits the crystal at the prismatic part of the needle near
the vertex has sticking probability of order unity, or that atoms
adsorbed in a ``wrong" place on the crystal somehow migrate towards
the end of the needle.

While most of the whiskers are perpendicular to the cell wall,
occasionally, we find needles that grow at an angle.

A quick e-mail poll of the workers who have studied paraffin coated
cells has revealed that, while most of them have not seen
alkali-metal whiskers, the paraffin-coating pioneer, Hugh Robinson
(now at NIST, Boulder), has, in fact, first observed their formation
some 20 years ago, and was even able to stimulate their formation by
providing a ``cold spot'' on the surface of the cell. Unfortunately,
Robinson's results remain unpublished.

As it turns out, rubidium is a member of a large ``family'' of
metals and other materials that tend to grow whiskers \cite{Giv87}.
In fact, tin, zinc, and gold whiskers represent a significant
problem for electronics industry, as they can lead to electrical
shorts and failure of electronic equipment. NASA maintains an
informative web site \cite{NASAWhisker}
devoted to metal whiskers, however, alkali metals have not as yet
been featured. There are also reports of bismuth whiskers growing in
a heat-pipe vapor cell \cite{Zol2006}.
\begin{figure}[h]
\includegraphics[width=3.2 in]{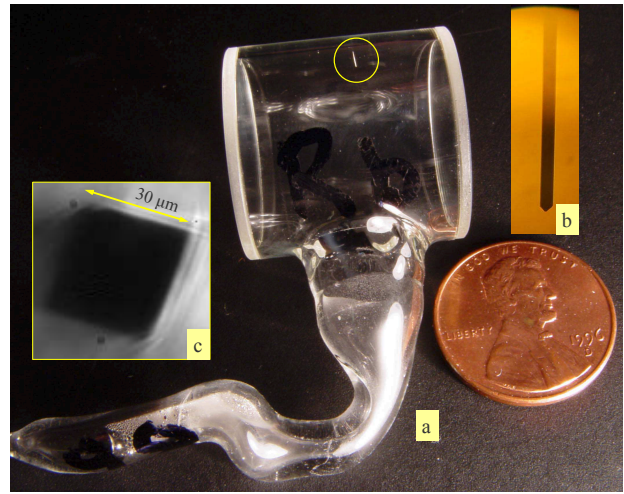}
\caption{a. The view of a paraffin coated vapor cell with one of the
rubidium whiskers clearly seen (outlined with a yellow circle). b. A
microscope photograph of the side view of the whisker. c. Microscope
photograph of the base of the whisker as seen from the coating side.
}\label{Fig_General_Sch}
\end{figure}

Alkali-metal needles may also present certain problems, for example,
in the work that requires application of high-voltage electric field
to coated cells \cite{Bud2002Bennett}. On the other hand, they might
also prove useful as a way to manufacture alkali-metal wires, for
which one interesting application is loading a dense cryogenic
buffer gas with the alkali metal by laser ablation from of the end
of a micron-sized wires \cite{Sus2006}.

We are grateful to  W. Happer, L. R. Hunter, H. G. Robinson, and M.
Zolotorev for useful comments, and to A. Cing\"{o}z and D. English
for help with photography.

\bibliography{NMObibl}

\end{document}